\documentclass[aps,pre,groupedaddress,showpacs,preprintnumbers,preprint]{revtex4}

\usepackage{graphicx}
\usepackage{color}
\usepackage[ansinew]{inputenc}
\usepackage{amsmath}
\usepackage{amsfonts}
\usepackage{amssymb}
\usepackage{dcolumn}
\usepackage{float}

\begin{document}

\title{Random Sequential Adsorption of Objects of Decreasing Size}

  \author{Oleksandr Gromenko}
  \affiliation{Department of Physics, Clarkson University, Potsdam, NY 13699--5721, USA}
  \author{Vladimir Privman}
  \affiliation{Department of Physics, Clarkson University, Potsdam, NY 13699--5721, USA}

  \pacs{68.43.Mn,\ 02.50.--r,\ 05.10.Ln,\ 05.70.Ln}

  \begin{abstract}
We consider the model of random sequential adsorption, with depositing objects,
as well as those already at the surface, decreasing in size according to a
specified time dependence, from a larger initial value to a finite value in the
large time limit. Numerical Monte Carlo simulations of two-dimensional deposition
of disks and one-dimensional deposition of segments are reported for the
density-density correlation function and gap-size distribution function,
respectively. Analytical considerations supplement numerical results in the
one-dimensional case. We investigate the {\it correlation hole\/} --- the depletion of correlation
functions near contact and, for the present model, their vanishing at contact --- that opens
up at finite times, as well as its closing and reemergence of the
logarithmic divergence of correlation properties at contact in the large
time limit.
  \end{abstract}

\maketitle

\section{Introduction}
Random sequential adsorption (RSA) model has attracted a lot of
attention and has a long history
\cite{Evans}, \cite{Privman1}, \cite{Privman3}, \cite{Privman55}.
It finds applications, e.g.,
\cite{Evans}, \cite{Privman3}, \cite{Privman55}, \cite{Ramsden}, \cite{Tassel}, \cite{Torquato1}, \cite{Luryi}, \cite{Privman2} in
many fields, ranging from surface science to polymers, biology, device physics, and
physical chemistry. Traditionally, the RSA model assumed that
particles are transported to a substrate which is a continuous
surface or a lattice, the latter convenient for numerical
simulations. Upon arrival at the surface the particles are
irreversibly deposited, but only provided they do not overlap
previously deposited objects. Otherwise, the deposition attempt is
rejected and the arriving particle is assumed  transported away
from the substrate. The original RSA model studies were largely
motivated by surface deposition of micron-size objects, such as
colloid particles. Particles of this size are typically not
equilibrated on the surface, and are larger than most surface features
and the range of most of the particle-particle and particle-surface
interactions. Various generalizations of the basic RSA model have
been considered in the literature, e.g.,
\cite{Luryi}, \cite{Roosbroeck}, \cite{Aling}, \cite{Inoue}, \cite{Viot}, \cite{Wang}, \cite{Boyer}, \cite{Adamszyk}, \cite{Rodgers}, \cite{Hassan}, \cite{Burridge}, \cite{Cadilhe}, \cite{Nielaba}, \cite{Wang2}, \cite{Privman8}, \cite{Privman5}, \cite{Privman6}, \cite{Nielaba2}, \cite{Privman7}, \cite{Privman9},
including ``soft'' rather than hard-core particle-particle
interactions, as well as relaxation by motion of particles
on the surface.

Recently, experimental surface-deposition work has expanded to
nano-size particles and sub-micron-feature patterned (ultimately,
nano-patterned) surfaces
\cite{Dziomkina}, \cite{Liddlea}, \cite{Ogawa}, \cite{Deshmukh}. The
added control of the particle and surface ``preparation'' as part
of the deposition process could allow new functionalities in
applications, and therefore it has prompted new research efforts.
Specifically, deposition on surfaces prepared with patterns other
than regular lattices was studied \cite{Privman5}, \cite{Privman6} motivated by new
experimental capabilities in surface patterning. Another
development involved a study \cite{Luryi}, \cite{Boyer}, \cite{Rodgers} of
one-dimensional deposition of segments that, after attachment to
the substrate, can shrink or expand, motivated by
potential applications, e.g., in device
physics \cite{Roosbroeck}, \cite{Aling}.

The motivation for our present work has been a newly
emerging experimental capability \cite{Robert}, \cite{MMotornov}
of depositing polymer ``blobs'' (or polymer-coated particles) the
size of which can be modified by changing the solution chemistry:
Approximately spherical particles can be deposited in a
process whereby their effective size (including the interaction
radius) is varied on a time scale comparable to that of the
deposit formation. The size of both the particles in
solution and those already deposited will thus vary with time, in
a controllable fashion.

For particle deposits formed on
pre-patterned surfaces, an interesting property is suggested by experiments,
\cite{Nath}, \cite{Tokareva1}, \cite{Zhong}, \cite{Lee}, \cite{Tokareva2}, \cite{Azzaroni}, \cite{Ding},
and theoretically verified, \cite{Privman5}, \cite{Privman6}:
They acquire semi-ordering properties ``imprinted'' by the
substrate, as quantified by the development of peaks in
two-particle correlation. However, as in the original
irreversible RSA model, there remains a significant peak at particle-particle
contact, which for deposition continuing indefinitely, at infinite times
becomes a weak singularity \cite{Pomeau}, \cite{Swendsen}. This tendency of particles
to form clumps, is undesirable mostly because many
nanotechnology applications rely on nanoparticles utilized in
isolation, or simply being kept away from each other to avoid merging.
In this work, we establish that deposition of particles of
varying, specifically, {\it decreasing\/} size, can yield deposits
without clumping, by opening a ``correlation hole,'' i.e., a property of
depletion of two-particle correlations near
contact, as defined in other fields, e.g., \cite{CH3}, \cite{CH4}, \cite{CH5}, \cite{CH1}, \cite{CH6}, \cite{CH2}. In our case,
the correlation functions studied actually vanish at contact for finite deposition times.

The outline of this article is as follows. In Section \ref{2D},
we consider the two-dimensional (2D) deposition of shrinking disks in a plane.
The model is defined and then a two-point density-density correlation
function is studied by numerical Monte
Carlo simulations. In Section \ref{1D}, we present both numerical and
analytical analysis of the one-dimensional (1D) deposition of shrinking
segments on a line. Our 1D study focuses on the gap-density distribution
function. A brief summary is offered at the end of Section \ref{1D}.

\section{Deposition of disks on a two-dimensional substrate}
\label{2D}

\subsection{Definition of the model}

In this section, we introduce the model for the most relevant
geometry for possible applications: The 2D RSA
model of deposition of disks of diameters $D(t)$ on an initially
empty planar substrate. A 1D model of deposition
of segments on a line provides additional insight into the problem
and will be considered in the next section.

As usual in RSA, we assume that disks are transported to the
surface with the resulting deposition attempt flux $R$, per unit
area and unit time, $t$. A disk adsorbs only if it does not
overlap a previously deposited one. The diameter of all the disks,
those already on the surface and those arriving, is a decreasing
function of time, $\dot D(t)<0$, varying between two nonzero
values $D(0) > D(\infty)>0$. For the 2D model, we carried out
numerical Monte Carlo simulations to estimate the
particle-particle pair correlation function that describes the
relative positioning of the disks with respect to each other. It
is defined as the ratio of the number of particle centers $N$ at
distances from $r$ to $r+dr$ from a given particle's center, normalized
per unit area and per the deposit density, $\rho$,
\begin{equation}
\label{definitionP} P_{2}(r,t)=\frac{N(r,dr,t)}{(2\pi rdr)\rho(t) }\, .
\end{equation}

In studies of RSA, $\rho(t)$ is usually a quantity of interest.
It grows linearly for short times and reaches a jamming-limit
value, which is less than close packing, for large times. In the
latter regime the approach to the jamming coverage on a continuum substrate is described by
a power law \cite{Pomeau}, \cite{Swendsen}, \cite{Feder}. However,
in our present study we found by numerical simulations no new
interesting features for $\rho(t)$, for the considered time-dependences of the disk
diameters (specified below). Therefore, we focus our presentation on the correlation
function, and specifically its properties near particle contact,
at $r \simeq D(t)$.

\begin{figure}[t]
\begin{center}
\includegraphics[width=4.0 true in]{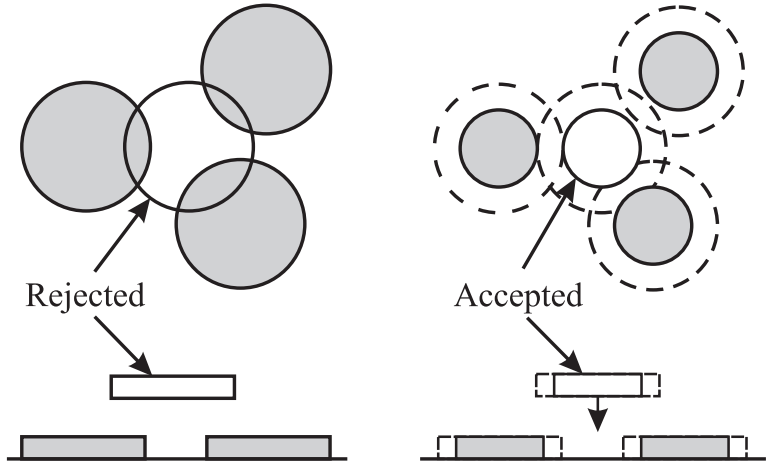}
\caption{The left panel illustrates deposition attempts in 2D and
1D that are rejected due to particle overlap. The right panel
shows a situation at a later time at which the same deposition
attempts would succeed due to particle shrinkage.}\label{Fig1}
\end{center}\hrule
\end{figure}

The model is illustrated in Figure\ \ref{Fig1}. We note that if the
disks shrink too fast as compared to the time scale of the buildup
of the deposited layer, which is of order $\sim 1/RD^2$, then the
problem will be reduced to that of simply depositing smaller
disks. On the other hand, if the disks shrink too slow as compared
to the time scale $\sim 1/RD^2$, then the depletion of the
correlations at contact will be trivially attributable to disk
shrinkage alone. We are interested in the interplay of two
effects: The decreasing disk size enlarges voids between already
deposited disks. At the same time this process increases the rate of
successful disk deposition events, which reduce voids between disks. Given that numerical
simulations for this problem are quite demanding, we report
results for the following time-dependence for the disk radius,
\begin{equation}
D(t) = D(\infty) \Big\{1+\exp[-RD^2(\infty)t]\Big\} \,.\label{dt1}
\end{equation}
We also studied numerically the cases
\begin{equation}
\label{dt2} D(t) = D(\infty) \Big\{1+\frac{1}{\ln[e+RD^2(\infty)t]}\Big\}\, ,
\end{equation}
\begin{equation}
\label{dt3} D(t) = D(\infty) \Big\{1+\frac{1}{1+RD^2(\infty)t}\Big\}\, ,
\end{equation}
and found qualitatively similar results.

\subsection{Numerical results}

Let us now briefly outline the numerical procedure used in our
Monte Carlo simulations. In order to calculate the correlation
function $P_{2}(r,t)$, we have to generate a distribution of
deposited objects. We used periodic boundary conditions and system
sizes $L \times L = 500D(\infty) \times 500D(\infty)$. The $xy$-coordinates of
the center of the next disc which makes a deposition  attempt were
randomly generated. If the disk does not overlap any of the
previously deposited ones, then it is deposited. The total number
of deposition attempts is thus equal to $R L^2 \Delta t$ for a
physical time interval $\Delta t$. All our data presented
below, were averaged over 100 independent Monte Carlo runs.

We point out that there exist algorithms for studying RSA at large
times \cite{Wang3}, specifically designed to account for the fact
that isolated residual voids (defined as landing areas for particle centers
for allowed depositions) then evolve independently. However, in our case we
were primarily interested in intermediate times. Furthermore, as
disks already adsorbed on the surface shrink, all the voids increase and some
might merge to form larger voids. Therefore, we used the
straightforward algorithm described in the preceding paragraph.

\begin{figure}[t]
\begin{center}
\includegraphics[width=4.0 true in]{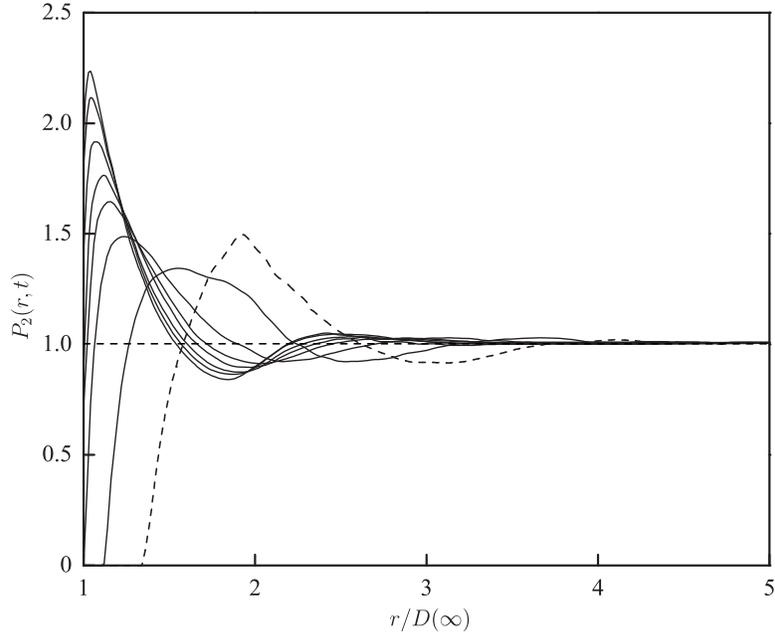}
\caption{Time evolution of the correlation function in deposition
of disks with diameters shrinking according to Eq.\ (\ref{dt1}). The
dashed line corresponds to $t=\tau$. The solid lines with
increasing peak values at $r/D(\infty)$ approaching 1, correspond,
respectively, to $t=2\tau$, $4\tau$, $6\tau$, $8\tau$, $12\tau$,
$18\tau$, $24\tau$.}\label{Fig2}
\end{center}\hrule
\end{figure}

Our main result is illustrated in Figure\ \ref{Fig2}. The
correlation function introduced earlier, is plotted for several
times, which are multiples of a characteristic process time-scale
\begin{equation}
\tau = 1/RD^2(\infty)\, .\label{tau}
\end{equation}
At finite times the correlation function has a peak at a value
$r/D(\infty)>1$, and actually vanishes at the disk contact, when
$r=D(t)$. This ``correlation hole'' behavior is in contrast with
the ordinary RSA, e.g., \cite{Torquato1}, \cite{Feder}, for which the correlation
function instead of developing a peak close to contact, actually increases as
the disk separation decreases towards contact, and reaches, for finite times, a
constant, non-zero value at contact. The ``correlation hole''
property represents a formation of a certain degree of short-range
ordering in the system, and specifically, avoidance of particle
clumping.

\begin{figure}[t]
\begin{center}
\includegraphics[width=4.0 true in]{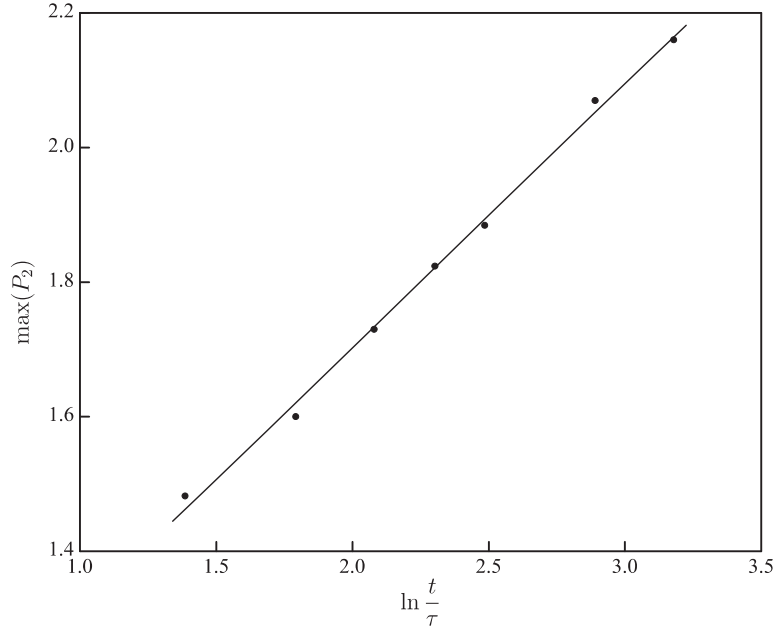}
\caption{Time-dependence of the peak values of the correlation
function, demonstrating the expected logarithmic divergence as $t
\to \infty$. The least-squares fit of the peak values, shown
here for times $t=4\tau$, $6\tau$, $8\tau$, $10\tau$, $12\tau$,
$18\tau$, $24\tau$, yields
the slope of the solid line as $0.37 \pm 0.02$ (but note a comment in Subsection\ \ref{subsecti} below regarding low reliability of this ``error-bar'' estimate).}\label{Fig3}
\end{center}\hrule
\end{figure}

In the limit $t \to \infty$, the correlation hole closes and the
correlation function seems to develop a weak divergence as $r \to
D(\infty)$. In fact, for constant disk diameters, this property has
been studied by asymptotic analytical arguments
\cite{Swendsen} and numerically \cite{Torquato1}, \cite{Feder}: The weak
logarithmic divergence of the correlation function of RSA at
contact, is not easy to quantify numerically, and our data were
not accurate enough to study this limit for shrinking-diameter
disk deposition by directly estimating $P_{2}(r,\infty)$. However,
the time-dependence of the peak values in Figure\ \ref{Fig2},
follows a logarithmic divergence as $t$ increases, as shown in
Figure\ \ref{Fig3}. This is reminiscent of a similar logarithmic
divergence predicted \cite{Swendsen} for the values of $P_{2}(D,t)$ at contact
($r=D$) for fixed-$D$ RSA.

\begin{figure}[t]
\begin{center}
\includegraphics[width=4.0 true in]{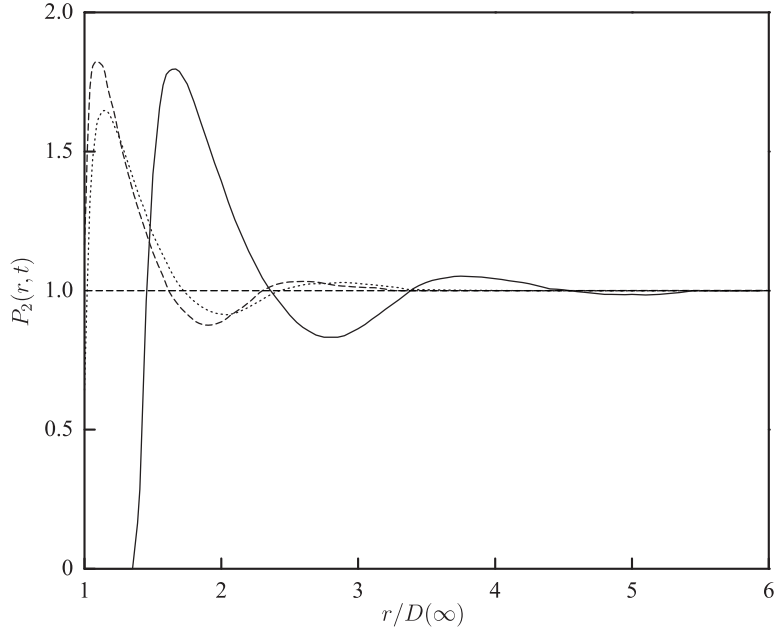}
\caption{The correlation function for $t=10\tau$, in deposition
of disks with diameters shrinking according to Eq.\ (\ref{dt1}): dashed line;
 Eq.\ (\ref{dt2}): solid line;  and Eq.\ (\ref{dt3}): dotted line.}\label{Fig4}
\end{center}\hrule
\end{figure}

The formation of the finite-time correlation hole at disk-contact,
was observed also for the non-exponential disk-diameter
time-dependences defined in Eqs.\ (\ref{dt2}-\ref{dt3}); see
Figure\ \ref{Fig4}. This property of the correlation function is
qualitatively similar for all three time-dependence protocols
studied. However, we did not study larger time values for the
non-exponential time-dependences.

\section{Deposition of Segments on a Line}
\label{1D}

\subsection{Numerical results}

In this section we report numerical results for
the 1D deposition of segments on
an initially empty line. The segment length is a function
of time, $\ell(t)$, monotonically decreasing from
$\ell(0)$ to $\ell(\infty)$. Our numerical results in 1D were obtained
for the exponential time
dependence similar to the 2D case,
\begin{equation}
\ell(t)=\ell(\infty)\Big\{1+\exp[-R\ell(\infty)t]\Big\} \,.\label{label}
\end{equation}
We denote the number of deposition attempts per unit time per
unit length (the flux) by $R$. The arriving segments are adsorbed only if they do not
overlap any previously deposited ones; see Figure\ \ref{Fig1}. The quantity of
interest is the gap density
distribution function, $G(x,t)$: The density of gaps (measured between the ends
of the nearest-neighbor deposited segments) of
length between $x$ and $x+dx$ at time $t$ is $G(x,t)dx$. The density of deposited
segments at time $t$, which in 1D equals the density of gaps, is given by
\begin{equation}
\label{Norm} n(t) = \int^{\infty}_{0}G(x,t)dx \,.
\end{equation}

\begin{figure}[t]
\begin{center}
\includegraphics[width=4.0 true in]{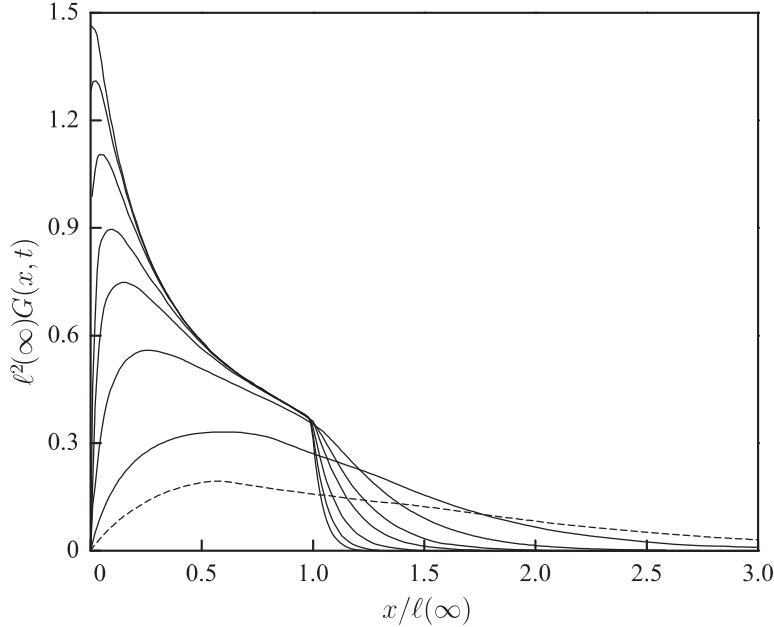}\caption{Time evolution of the dimensionless gap-size distribution function $\ell^2(\infty)G(x,t)$ for segments with length shrinking according to Eq.\ (\ref{label}). The
dashed line corresponds to $t=\tau$. The solid lines with
increasing peak values at $x/\ell(\infty)$ eventually approaching 0, correspond,
respectively, to $t=2\tau$, $4\tau$, $6\tau$, $8\tau$, $12\tau$,
$18\tau$, $24\tau$.}\label{Fig5}
\end{center}\hrule
\end{figure}

Our numerical simulations followed the same procedure as in 2D. In 1D, we used
periodic boundary conditions and system sizes $10000\ell(\infty)$.
The time scale, cf.\ Eq.\ (\ref{tau}), was redefined for 1D,
\begin{equation}
\tau \equiv 1/R\ell(\infty) \,.
\end{equation}
Our results, shown in Figure\ \ref{Fig5}, represent averages over 10000
independent Monte Carlo runs.
One can see that the gap density distribution
function is a smooth function of $x$ for finite times, $t$, and for larger times
it develops a pronounced peak at $x<\ell(t)$ approaching $x=0$. However, at $x=0$ for finite times there is a ``correlation hole.'' In the large-time limit,
$G(x,t\to\infty)$ appears to develop
a step at $x=\ell(\infty)$, as well as a divergence at $x=0$. These properties will be further discussed in the next two subsections.

\subsection{Analytical considerations}

In this subsection, we will consider an analytical kinetic
equation approach to the 1D deposition problem. This will allow us
to elucidate the origin of the ``correlation hole'' at $x=0$.
Kinetic equations, closed-form or hierarchies of, for RSA can be
formulated for the gap density distribution function as well as
for several other correlation-function-type quantities. They have
been widely used for deriving exact results and low-density
approximation schemes; see reviews in \cite{Evans}, \cite{Privman1}, \cite{Privman3}, \cite{Privman55}.
For the present problem, we have
\begin{equation}
\label{Initial_equation} \frac{\partial G(x,t)}{\partial t} =
-R\theta
(x-\ell(t))[x-\ell(t)]G(x,t)+2R\int^{\infty}_{x+\ell(t)}G(y,t)dy +
\dot{\ell}(t)\frac{\partial G(x,t)}{\partial x}\, .
\end{equation}
Here the first term represents the destruction of gaps of length
$x$ by segment deposition, with $\theta (\ldots)$ denoting the
standard step function which vanishes for negative arguments and
is 1 for positive arguments. The second term represents the
creation of gaps of length $x$ due to deposition of segments landing in larger gaps of size $y > x + \ell(t)$.
The last term describes the increasing size of all the gaps as a result
of segment shrinkage.
Note that the case $\dot \ell (t) = 0$ is exactly solvable
\cite{Gonzalez}, and we will be using some of the exact results in the next subsection.
For large $x$, we obviously have the boundary condition $G(x \to \infty, t>0) = 0$ for Eq.\ (\ref{Initial_equation}).
The following discussion explores the form of the boundary condition at the small-$x$ side of the distribution.

There are two kinetic processes that alter gap lengths in our problem. First, deposition of segments
can reduce the length by $\ell(t)$. This process can generate very small, positive gap lengths, $\epsilon$,
from available gaps of sizes $\ell(t) + \epsilon$. At the same time, the second process, that of segment
shrinkage, causes all gap lengths to ``drift'' towards larger positive values with velocity $-\dot \ell (t)$.
These two kinetic processes determine the evolution of the gap-length distribution $G(x,t)$ from its
initial value $G(x,0)$. While our numerical simulations were for $G(x,0)=0$, we note that in principle one
can generate other translationally invariant initial distributions by an appropriate preparation of the initial state. In fact, one
can even prepare initial distributions that extend to small {\it negative\/} gap lengths. This entails allowing
overlaps for the segments initially placed on the line, but not for those later depositing. If the segments are
ordered according to their center-point positions, and each gap is measured as a consecutive center-point distance minus $\ell (t)$,
then the initial-distribution overlaps, even if some are multi-segment, can be unambiguously counted as negative gaps.
The two kinetic processes will then remain the same: Gaps larger than $\ell (t)$ can be shortened
due to deposition events, while at the same time {\it all\/} the gaps (positive, zero, negative) also increase towards positive
values, due to segment shrinkage.

The above considerations suggest that equation Eq.\ (\ref{Initial_equation}) strictly speaking should be mathematically considered for
$-\infty < x < \infty$. An attempt to limit it to $0 \le x < \infty$, and also use moment definitions, such as the zeroth-order
moment Eq.\ (\ref{Norm}), with integration over $0 \le x < \infty$, may in general yield wrong results: Neglecting the ``flow of length'' from the
negative-$x$ values may violate length conservation. However, in our case, specifically for the initial
condition $G(x,0)=0$ (and with nonvanishing $\ell (t)$ for all $t$, including in the $t \to \infty$ limit, which avoids a possible
singular limit), it is obvious that the problem should be definable with a boundary condition at $x=0$ because all the kinetics of the
process occurs in $0 \le x < \infty$. Indeed, together with the
zeroth-order moment, the first-order moment of the distribution can be used to determine the applicable boundary condition
directly from length-conservation,
\begin{equation}
\ell(t) n(t) + \int_0^\infty x G(x,t)dx = 1 \, .
\end{equation}
Here the first term is the density of the
covered area, whereas the second term is the density of the uncovered area.
Both densities are ``per unit length'' in 1D, and therefore they sum up to 1.
Differentiation with respect to time, $t$, and the use of Eq.\ (\ref{Initial_equation}),
then yield, after some algebra, the result
\begin{equation}
\dot{\ell}(t)G(0,t)=0 \, . \label{BCO}
\end{equation}
For shrinking segments, $\dot{\ell} < 0$, we are thus led to our main conclusion,
\begin{equation}
G(0,t)=0 \, ,
\end{equation}
except perhaps in the limit $t \to \infty$ (in which $\dot \ell$ vanishes). This result indicates that the
``correlation hole'' for finite times, at $x=0$ is generic for the shrinking-segment RSA, no matter how fast is the
deposition kinetics that tends to ``fill the hole'' in the distribution at small $x$: The presence
of the correlation hole (the depletion of the gap-distribution near $x=0$) follows from the boundary
condition established for $x=0$.

\subsection{Properties of the gap-length distribution}\label{subsecti}

\begin{figure}[t]
\begin{center}
\includegraphics[width=4.0 true in]{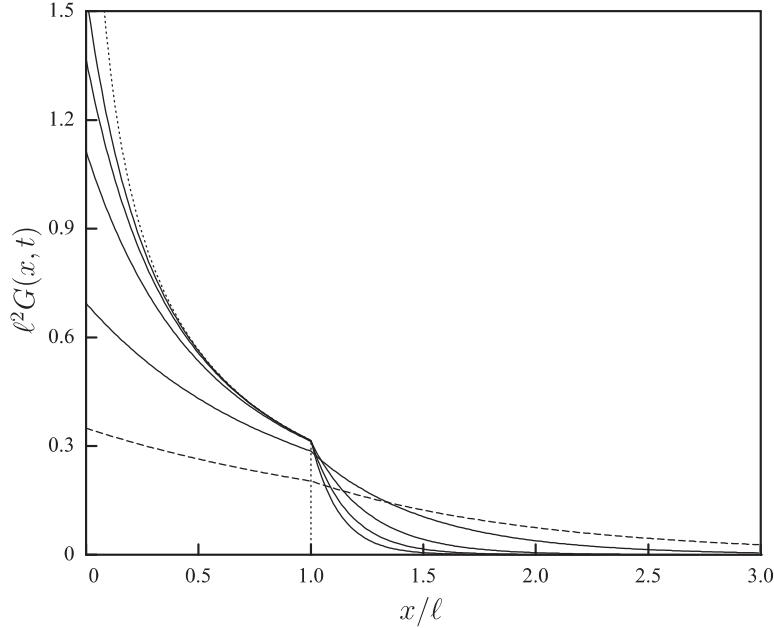}\caption{Time evolution of the dimensionless gap-size distribution function $\ell^2 G(x,t)$ for
segments of constant length $\ell$. The
dashed line corresponds to $t=\tau$. The solid lines correspond to $t=2\tau$, $4\tau$, $6\tau$, $8\tau$, with
increasing peak values at $x=0$, increasing values at $x = \ell$, and with respectively smaller values for $x\gg \ell$. The dotted line corresponds to $t \to \infty$,
with the values for $x>\ell$ equal 0 in this limit.}\label{Fig6}
\end{center}\hrule
\end{figure}

For ordinary 1D RSA of segments of fixed length $\ell(t) =\ell \equiv \ell(\infty)$, Figure\ \ref{Fig6} illustrates the
exact solution \cite{Gonzalez} for several times and in the limit $t\to \infty$. Specifically, the gap distribution has a finite
value at $x=0$, consistent with Eq.\ (\ref{BCO}) for $\dot \ell =0$. This value actually diverges in the $t \to \infty$ limit,
while a logarithmic singularity develops near $x=0$.
This suggests that ordinary RSA is actually a rather singular limit of the more
general shrinking-segment RSA.
Indeed, the point $x=\ell(t)$ corresponds to discontinuity in the first, deposition term in the
kinetic equation, Eq.\ (\ref{Initial_equation}). As a result, the fixed-$\ell$ RSA gap-distribution
has a discontinuous derivative at $x=\ell$, which becomes an actual discontinuity (jump) in the
$t \to \infty$ limit. On the
other hand, with segment shrinking allowed the distribution is apparently continuous and
smooth (as numerically observed), likely analytic
at all points internal to the domain of definition, $0<x<\infty$, of the gap distribution.
Adding the segment shrinkage process seems to smooth the singularities out for finite times;
cf.\ Figure\ \ref{Fig5}. In terms of the kinetic
equation, this is a consequence of that the added third, $\sim \partial G / \partial x$ term plays the
role of the ``diffusive'' (second-derivative) smoothing contribution with respect to the $x$-dependent
integral (the second term).
This diffusive property of the kinetic equation suggests that the function $G(x,t)$ has no singularities
for any $x>0$ for finite time, though in the limit $t \to \infty$ (when $\dot \ell (t) \to 0$) the
divergence as $x \to 0^+$ and the discontinuity at $x = \ell (\infty)$ are asymptotically restored.

The exact solution technique for the constant-length case \cite{Gonzalez} involves the use of the
exponential-in-$x$ ansatz for $G(x,t)$ for $x > \ell$, recently detailed in applications for
related models in \cite{BN}, \cite{Privman7}, and then
solution of the $x <\ell$ equation, Eq.\ (\ref{Initial_equation}),
which becomes tractable because the first term is not present,
whereas the integration in the second term involves a simple
exponential integrand. Thus, this approach by its nature yields
discontinuities if not in the function then in its first or
possibly higher-order derivatives. Attempts to use this approach,
as well as more complicated, exponential-multiplied-by-polynomial ansaetze
for $x > \ell(t)$, for time-dependent $\ell(t)$ yield solutions
which satisfy the equation but possess unphysical discontinuities.
We consider it unlikely that the shrinking-segment RSA problem in 1D can
be solved exactly by the presently known techniques.

\begin{figure}[t]
\begin{center}
\includegraphics[width=4.0 true in]{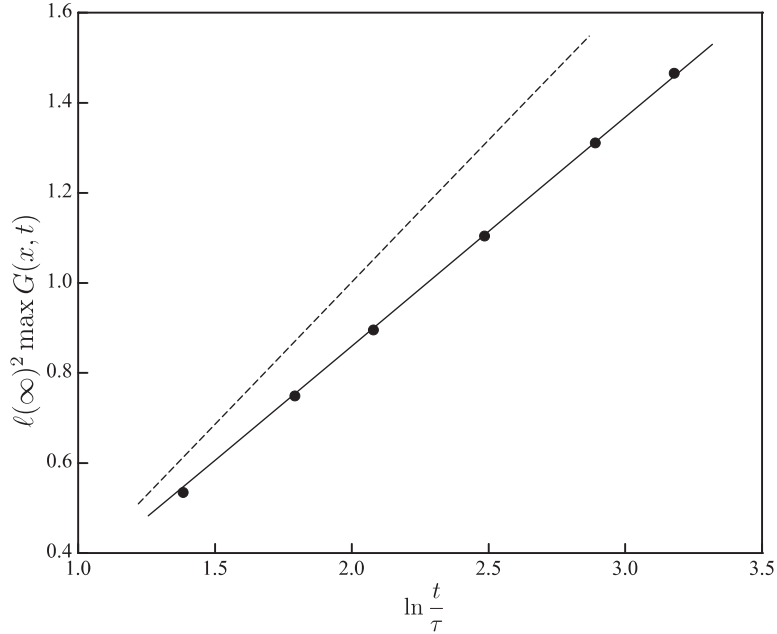}\caption{The dashed straight line is the asymptotic large-time form of the gap-distribution function for the fixed-segment-length RSA; its slope is $2e^{-2\gamma}$ see Eq.\ (\ref{Slope}). The solid straight line is a fit of our data (shown as symbols) for $t \geq 4 \tau$, for shrinking-segment RSA, with the estimated slope of $0.51 \pm 0.02$.}\label{Fig7}
\end{center}\hrule
\end{figure}

Finally, let us consider the extent to which our numerical data
can confirm the restoration of the logarithmic singularity near
$x=0$ in the $t \to \infty$ limit. As in 2D, our data are not accurate
enough to confirm the form of the $x$-dependence of the expected
logarithmic divergence for $x \ll \ell(\infty)$, i.e., $G(x ,\infty) \propto -\ln \left(x/\ell(\infty)\right)$.
Regarding the $t$-dependence, Figure\ \ref{Fig7} shows our
1D data for the peak values at small positive $x$, see
Figure\ \ref{Fig5}, as well as the asymptotic large-time form of
the exact solution peak values at
$x=0$ for constant $\ell = \ell(\infty)$, the latter given by the
relation \cite{Gonzalez}
\begin{equation}\label{Slope}
\ell^2 G(t,0) \simeq 2e^{-2\gamma}\ln ({t}/{\tau}), \qquad t \gg \tau \, ,
\end{equation}
where $\gamma=0.5772156649\ldots$ is Euler's gamma constant. Typical for
data fits for logarithmic dependences, the estimated slope given in
Figure\ \ref{Fig7}, may gradually drift as the
$t$-variable varies over many decades. It is possible that it reaches the larger
value of the fixed-segment RSA: We consider the error estimate,
0.02, here and in the 2D case (Figure\ \ref{Fig3}) inconclusive.

In summary, we investigated an RSA model in which deposition of arriving
objects, which tends to form small gaps/voids, competes with the process of shrinking
of the objects, both those already deposited and those newly arriving. The
latter process always ``wins'' at finite time, and as a result ``correlation
hole'' opens up in the correlation properties considered. In fact, analytical
considerations in 1D suggest that this property is generic for an initially
empty substrate. Numerical Monte Carlo
results reported, support this conclusion for the time-dependences studied.
For the exponential-decay time dependence of the object sizes studied in 1D and 2D,
we also found preliminary evidence that logarithmic divergence of the correlation
properties at contact is restored as $t \to \infty$. For 1D, the discontinuous behavior
of the gap-distribution at $x=\ell(\infty)$ also reemerges in this limit.

The authors thank Drs.\ S.\ Minko and I.\ Tokarev for instructive discussions,
and acknowledge support of this research by the NSF under grant DMR-0509104.


\begin{thebibliography}{00} {\frenchspacing

\bibitem{Evans}
Review:\ J.\ W.\ Evans,\ Rev.\ Mod.\ Phys.\ \textbf{65},\ 1281\ (1993).

\bibitem{Privman1}
Review:\ M.\ C.\ Bartelt\ and\ V.\ Privman,\ Int.\ J.\ Mod.\ Phys.\ B\ \textbf{5},\ 2883\ (1991).

\bibitem{Privman3}
Collection of review articles:\ {\it{}Adhesion\ of\ Submicron\ Particles\ on\ Solid\ Surfaces\/}, V.\ Privman,\ ed.,\ special\ volume\ of\ Colloids\
and\ Surfaces\ A\ \textbf{165},\ Issues\ 1-3,\ Pages\ 1-428\ (2000).

\bibitem{Privman55}
Collection of review articles:\ {\it{}Nonequilibrium\ Statistical\ Mechanics\ in\ One\ Dimension\/}, V.\ Privman,\ ed.\ (Cambridge University Press, Cambridge, 1997).

\bibitem{Ramsden}
Review: J.\ J.\ Ramsden,\ Chem.\ Soc.\ Rev.\ \textbf{24},\ 73\ (1995).

\bibitem{Tassel}
P.\ R.\ Van\ Tassel,\ P.\ Viot\ and\ G.\ Tarjus,\ J.\ Chem.\ Phys. \textbf{106},\ 761\ (1997).

\bibitem{Torquato1}
S.\ Torquato,\ O.\ U.\ Uche\ and\ F.\ H.\ Stillinger,\ Phys.\ Rev.\ E\ \textbf{74}, 061308 (2006).

\bibitem{Luryi}
A.\ V.\ Subashiev and S.\ Luryi, Phys.\ Rev.\ E \textbf{75},\ 011123\ (2007).

\bibitem{Privman2}
Review:\ V.\ Privman,\ J.\ Adhesion\ \textbf{74},\ 421\ (2000).

\bibitem{Roosbroeck}
W.\ van\ Roosbroeck,\ Phys.\ Rev.\ \textbf{139},\ A1702\ (1965).

\bibitem{Aling}
R.\ C.\ Alig,\ S.\ Bloom\ and\ C.\ W.\ Struck,\ Phys.\ Rev.\ B\ \textbf{22}, 5565 (1980).

\bibitem{Inoue}
M.\ Inoue,\ Phys.\ Rev.\ B\ \textbf{25},\ 3856\ (1982).

\bibitem{Viot}
P.\ Viot,\ G.\ Tarjus\ and\ J.\ Talbot,\ Phys.\ Rev.\ E\
\textbf{48},\ 480\ (1993).

\bibitem{Wang}
J.-S.\ Wang,\ P.\ Nielaba\ and\ V.\ Privman,\ Physica\ A\ \textbf{199},\ 527-538\
 (1993).

\bibitem{Boyer}
D.\ Boyer,\ J.\ Talbot,\ G.\ Tarjus,\ P.\ Van\ Tassel\ and\ P.\ Viot,\ Phys.\ Rev.\ E\ \textbf{49}, 5525 (1994).

\bibitem{Adamszyk}
Z.\ Adamczyk\ and\ P.\ Warszy\'{n}ski,\ Adv.\ Colloid\
Interf.\ Sci.\ \textbf{63}, 41\ (1996).

\bibitem{Rodgers}
G.\ J.\ Rodgers\ and\ Z.\ Tavassoli,\ Phys.\ Lett.\ A\ \textbf{246}, 252 (1998).

\bibitem{Hassan}
M.\ K.\ Hassan,\ J.\ Schmidt,\ B.\ Blasius\ and\ J.\ Kurths,\ Phys.\ Rev.\ E \textbf{65}, 045103R (2002).

\bibitem{Burridge}
D.\ J.\ Burridge\ and\ Y.\ Mao,\ Phys.\ Rev.\ E\ \textbf{69}, 037102 (2004).

\bibitem{Cadilhe}
N.\ A.\ M.\ Ara\'{u}jo\ and\ A.\ Cadilhe,\ Phys.\ Rev.\ E\ \textbf{73}, 051602\ (2006) .

\bibitem{Nielaba}
P.\ Nielaba\ and\ V.\ Privman,\ Modern\ Phys.\ Lett.\ B
\textbf{6}, 533 (1992).

\bibitem{Wang2}
J.-S.\ Wang,\ P.\ Nielaba\ and\ V.\ Privman,\ Modern\ Phys.\
Lett.\ \textbf{B}\ \textbf{7}, 189 (1993).

\bibitem{Privman8}
M.\ C.\ Bartelt and V.\ Privman, Phys.\ Rev.\ A \textbf{44}, R2227 (1991).

\bibitem{Privman5}
A.\ Cadilhe, N.\ A.\ M.\ Ara\'{u}jo and V.\ Privman, J.\ Phys.\ Cond.\ Matter \textbf{19}, 065124 (2007).

\bibitem{Privman6}
N.\ A.\ M.\ Ara\'{u}jo, A.\ Cadilhe and V.\ Privman, Phys.\ Rev.\ E \textbf{77}, 031603 (2008).

\bibitem{Nielaba2}
V.\ Privman\ and\ P.\ Nielaba,\ Europhys.\ Lett.\ \textbf{18},
673\ (1992).

\bibitem{BN}
E.\ Ben-Naim and P.\ L.\ Krapivsky, Phys.\ Rev.\ E \textbf{54}, 3562 (1996).

\bibitem{Privman7}
O.\ Gromenko, V.\ Privman and M.\ L.\ Glasser, J.\ Comput.\ Theor.\ Nanosci., in print (2008).

\bibitem{Privman9}
V.\ Privman, Europhys.\ Lett.\ \textbf{23}, 341 (1993).

\bibitem{Dziomkina}
N.\ V.\ Dziomkina\ and\ G.\ J.\ Vancso,\ Soft\ Matter\ \textbf{1},\ 265\ (2005).

\bibitem{Liddlea}
J.\ A.\ Liddle,\ Y.\ Cui\ and\ P.\ Alivisatos,\ J.\ Vac.\ Sci.\ Technol.\ B\ \textbf{22}, 3409 (2004).

\bibitem{Ogawa}
T.\ Ogawa,\ Y.\ Takahashi,\ H.\ Yang,\
K.\ Kimura,\ M.\ Sakurai\ and\ M.\ Takahashi,\ Nanotechnology\ \textbf{17},\ 5539\ (2006).

\bibitem{Deshmukh}
R.\ D.\ Deshmukh,\ G.\ A.\ Buxton,\ N.\ Clarke\ and\ R.\ J.\ Composto,\ Macromolecules \textbf{40},\ 6316\ (2007).

\bibitem{Robert}
R.\ Lupitskyy,\ M.\ Motornov\ and\ S.\ Minko,\ Langmuir\ \textbf{24},\ 8976\ (2008).

\bibitem{MMotornov}
M.\ Motornov,\ R.\ Sheparovych,\ R.\ Lupitskyy,\ E.\ MacWilliams\
and\ S.\ Minko,\ J.\ Colloid\ Interf.\ Sci.\
\textbf{310},\ 481\ (2007).

\bibitem{Nath}
N.\ Nath and\ A.\ J.\ Chilkoti,\ J.\ Am.\ Chem.\ Soc.\ \textbf{123},\ 8197\ (2001).

\bibitem{Tokareva1}
I.\ Tokareva,\ S.\ Minko,\ J.\ H.\ Fendler and\ E.\ Hutter,\ J.\ Am.\ Chem.\ Soc.\ \textbf{126},\ 15950 (2004).

\bibitem{Zhong}
Z.\ Y.\ Zhong,\ S.\ Patskovskyy,\ P.\ Bouvrette,\ J.\ H.\ T.\ Luong and A.\ Gedanken, J.\ Phys.\ Chem.\ B \textbf{108},\ 4046\ (2004).

\bibitem{Lee}
S.\ Lee and\ V.\ H.\ P\'erez-Luna,\ Anal.\ Chem.\ \textbf{77},\ 7204 (2005).

\bibitem{Tokareva2}
I.\ Tokareva,\ I.\ Tokarev,\ S.\ Minko,\ E.\ Hutter and J.\ H.\ Fendler,\ Chem.\ Commun.\ 3343\ (30 June 2006).

\bibitem{Azzaroni}
O.\ Azzaroni,\ A.\ A.\ Brown,\ N.\ Cheng,\ A.\ Wei,\ A.\ M.\ Jonas and W.\ T.\ S.\ Huck,\ J.\ Mater.\ Chem.\ \textbf{17}, 3433 (2007).

\bibitem{Ding}
Y.\ Ding,\ X.\ H.\ Xia and\ H.\ S.\ Zhai,\ Chem. Eur.\ J. \textbf{13}, 4197 (2007).

\bibitem{Pomeau}
Y.\ Pomeau, J.\ Phys.\ A \textbf{13},\ L193 (1980).

\bibitem{Swendsen}
R.\ H.\ Swendsen,\ Phys.\ Rev.\ A \textbf{24},\ 504\ (1981).

\bibitem{CH3}
A.\ Yethiraj, C.\ K.\ Hall and K.\ G.\ Honnell, J.\ Chem.\ Phys.\ \textbf{93}, 4453 (1990).

\bibitem{CH4}
Y.\ C.\ Chiew, J.\ Chem.\ Phys.\ \textbf{93}, 5067 (1990).

\bibitem{CH5}
E.\ Kierlik and M.\ L.\ Rosinberg, J.\ Chem.\ Phys.\ \textbf{99}, 3950 (1993).

\bibitem{CH1}
J.\ Huh, O.\ Ikkala and G.\ ten Brinke, Macromol.\ \textbf{30}, 1828 (1997).

\bibitem{CH6}
V.\ Krakoviack, E.\ Kierlik, M.-L.\ Rosinberg and G.\ Tarjus, J.\ Chem.\ Phys.\ \textbf{115}, 11289 (2001).

\bibitem{CH2}
{\it{}Strongly Coupled Coulomb Systems\/}, G.\ J.\ Kalman, J. M.\ Rommel and K.\ Blagoev, eds. (Springer, New York, 2002).

\bibitem{Feder}
J.\ Feder, J.\ Theor.\ Biol.\ \textbf{87},\ 237\ (1980).

\bibitem{Wang3}
J.-S.\ Wang,\ Int.\ J.\ Mod.\ Phys.\ C\ \textbf{5},\ 707\ (1994).

\bibitem{Gonzalez}
J.\ J.\ Gonzalez,\ P.\ C.\ Hemmer\ and\ J.\ S.\ H{\o}ye,\ Chem.\ Phys.\ \textbf{3},\ 288\ (1974).

}\end{thebibliography}
\end{document}